# Catastrophe Observation in a Josephson Junction System

M. G. Castellano[1], F. Chiarello[1], R. Leoni[1], F. Mattioli[1], G. Torrioli[1],  P. Carelli[2], M. Cirillo[3], C. Cosmelli[4], A. de Waard[5], G. Frossati[5],  N. Grønbech-Jensen[6], and S. Poletto[7]

[1]Istituto di Fotonica e Nanotecnologie del CNR and INFN, 00156 Roma, Italy

[2]Dipartimento di Ingegneria Elettrica and INFN, Università dell'Aquila, 67040 L'Aquila, Italy

[3]Dipartimento di Fisica and INFM, Università di Roma Tor Vergata, 00173 Roma, Italy

[4]Dipartimento di Fisica and INFN, Università di Roma La Sapienza, 00195 Roma, Italy

[5]Kamerlingh Onnes Laboratory, Leiden University, 2333 CA Leiden, The Netherlands

[6]Department of Applied Science, University of California, Davis CA 95616, USA

[7]Dipartimento di Fisica, Università di Roma Tre, 00146 Roma, Italy

## Abstract

We report on experiments performed to probe quantum coherence between the two potential wells of a double SQUID system in the absence of external rf-signals. The system consists essentially of an rf-SQUID in which the Josephson junction interrupting the superconducting loop is replaced by another (smaller) loop containing two junctions in parallel. Experimental evidence at temperatures of the order of 10 mK shows that the system may develop three potential energy wells, which modify the usual two well energy profile and thereby vanify the qubit manipulation strategy. Analysis shows that the appearance of the third potential well can be interpreted as evidence of a butterfly catastrophe, namely a catastrophe expected for a system described by four control parameters and one state variable. The experimental results are interpreted on the basis of projections of the folded behaviour surface in the planes of the experimental control parameters.

Since the pioneering work by Leggett and Garg[1], the observation of quantum behavior of macroscopic superconducting variables has renewed the attention toward Josephson systems and SQUIDs (Superconducting Quantum Interference Devices). The response and dynamics of systems consisting of single[2] or coupled[3] Josephson junctions and interferometers[4] have been proposed and investigated in order to understand the nature of fundamental states and transitions between them. The acquired knowledge has been exploited in the growing field of quantum information processing with solid state devices[5]. In this framework, we have engineered a system that requires no external microwave pumping in order to provide evidence of coherent behavior because it relies only on the tunable configurations of a two well potential.

We study the properties of the system whose electrical analogue is sketched in Fig. 1a. The planar configuration consists in essence of a double-SQUID, namely a superconducting loop with inductance L interrupted by a small dc-SQUID with inductance $l$. When the effect of the small inductance $l$ can be neglected the inner dc-SQUID can be viewed as a single Josephson junction with tunable critical current, and the potential energy of the system has the form of the corrugated parabola of an rf-SQUID. This potential can be tilted by the applied flux $\Phi_x$ (Fig. 1b) and manipulated through the flux $\Phi_c$, which can lower the barrier (Fig. 1c). The readout occurs through a dc-SQUID magnetometer or through a larger junction inserted in the loop, in both cases by ramping their respective bias current $I_b$ (both schemes are sketched in Fig. 1a).

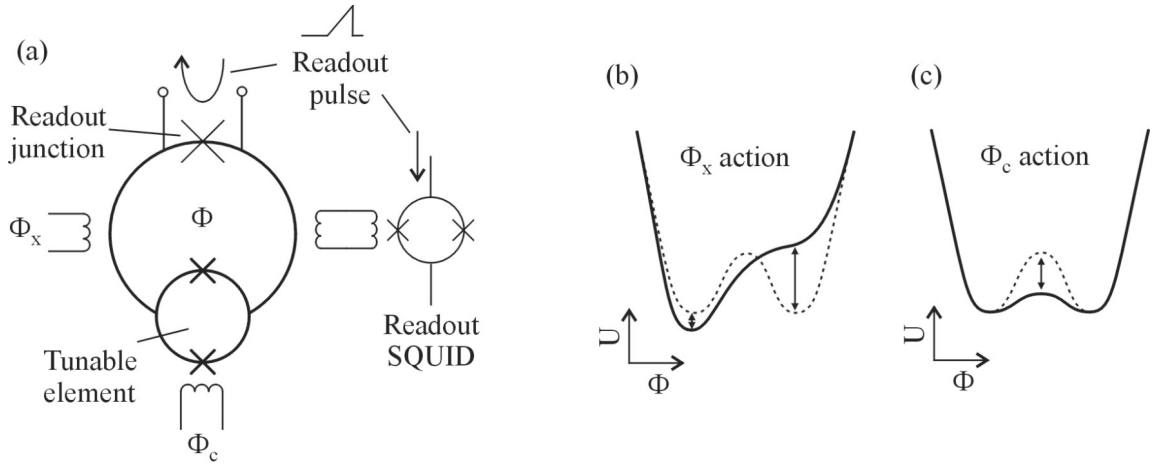

Figure 1. (a) The double SQUID system whose configurations and stability are investigated in the paper. The switching between flux states, induced by the fluxes $\Phi_x$ and $\Phi_c$, can be recorded through the readout junction or through the readout SQUID. Examples of the control of the system potential through $\Phi_x$ and $\Phi_c$ are given in (b) and (c), respectively: the first modifies the symmetry while the second tunes the height of the barrier.

The states in the right and left well of Fig. 1b,c correspond to clockwise and counter-clockwise current polarization states of the large loop. When the barrier of the potential well is very low and the temperature of the system is well below the expected quantum to classical crossover temperature[2] one can expect coherent oscillations between these two polarization states. These states can be measured and characterized either by a hysteretic dc-SQUID, coupled through a superconducting transformer, or by a larger Josephson junction, inserted into the double SQUID loop (Fig. 1a). In the first case, the double SQUID induces a magnetic flux in the readout dc-SQUID and hence a modulation of its critical current, whose value can be easily inferred from the measured switching current distribution[6]. In order to do so, the dc-SQUID current bias is ramped (with a repetition time of 100 μs) in such a way that during each cycle the critical current is exceeded and there is a switch from the zero voltage to the running state; the voltage discontinuity across the device triggers the acquisition board for reading the value of the switching current. The average of 100-1000 events allows us to infer a measure of the actual critical current. In the case of readout junction, the current circulating in the double SQUID adds to the bias ramp, so changing the amount of bias current needed to reach the critical value. Therefore, a measurement of the switching current distribution for the junction recovers the information of the loop current of the double SQUID.

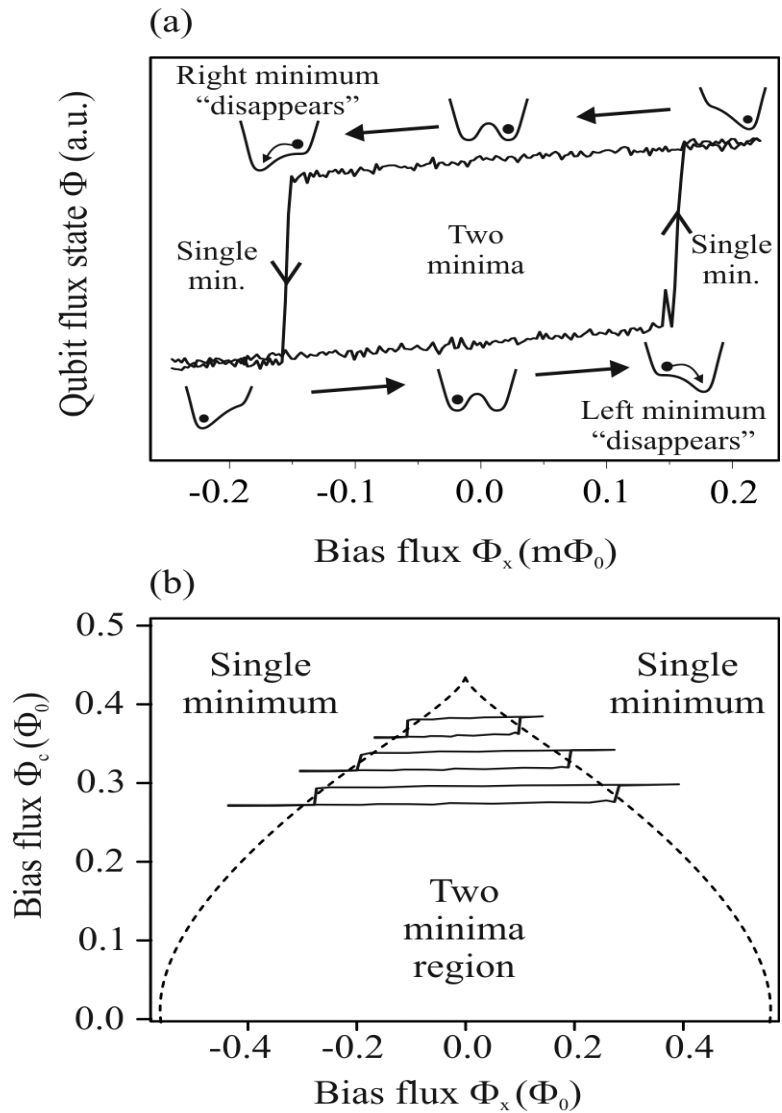

Figure 2. (a) An experimental current-voltage characteristics of the double SQUID system for a fixed value of $\Phi_c$ where we also report for every region of the curve the potential configuration. (b) A diagram of the potential configuration in the ($\Phi_x$, $\Phi_c$)-

plane, obtained by plotting the positions at which the two minima disappear. For clarity, we superimpose the characteristics shown in (a) taken for various values of $\Phi_c$.

In Fig. 2a we show a typical total flux vs control flux $\Phi_x$ characteristics, taken at 10 mK on our double SQUID with a fixed value of the control flux $\Phi_c$; in the figure we sketch for every portion of the characteristics the corresponding shape of the potential energy. We see that the characteristics are essentially those of an rf-SQUID[7], however, tuning the height of the potential barrier, which can be achieved just by varying the flux $\Phi_c$, will result in a reduction/enlargement of the hysteresis cycle. Recording the characteristics of Fig. 2a for different values of the flux $\Phi_c$ we will obtain the separatrix for the potential energy configurations in the ($\Phi_x$, $\Phi_c$)- plane as shown in Fig. 2b. Since the purpose of our experiment is to investigate possible quantum coherence of the SQUID states of the right and left well of the potential (with the interaction just regulated by $\Phi_c$) we found the plane of Fig. 2b to be a physically relevant and versatile tool for our investigations. In particular, the tip of the map is the most important region for our investigations because this is the region where the potential barrier between the two wells is the lowest, and, thus, where we can expect evidence of quantum coherence from the occupation statistics histograms.

In Fig. 3 we show the experimental data (open circles) corresponding to the map sketched in Fig. 2b. Figure 3a shows that the tip is bending on the left; the detail of the very top, indicated by the dotted square, is enlarged in Fig. 3b.

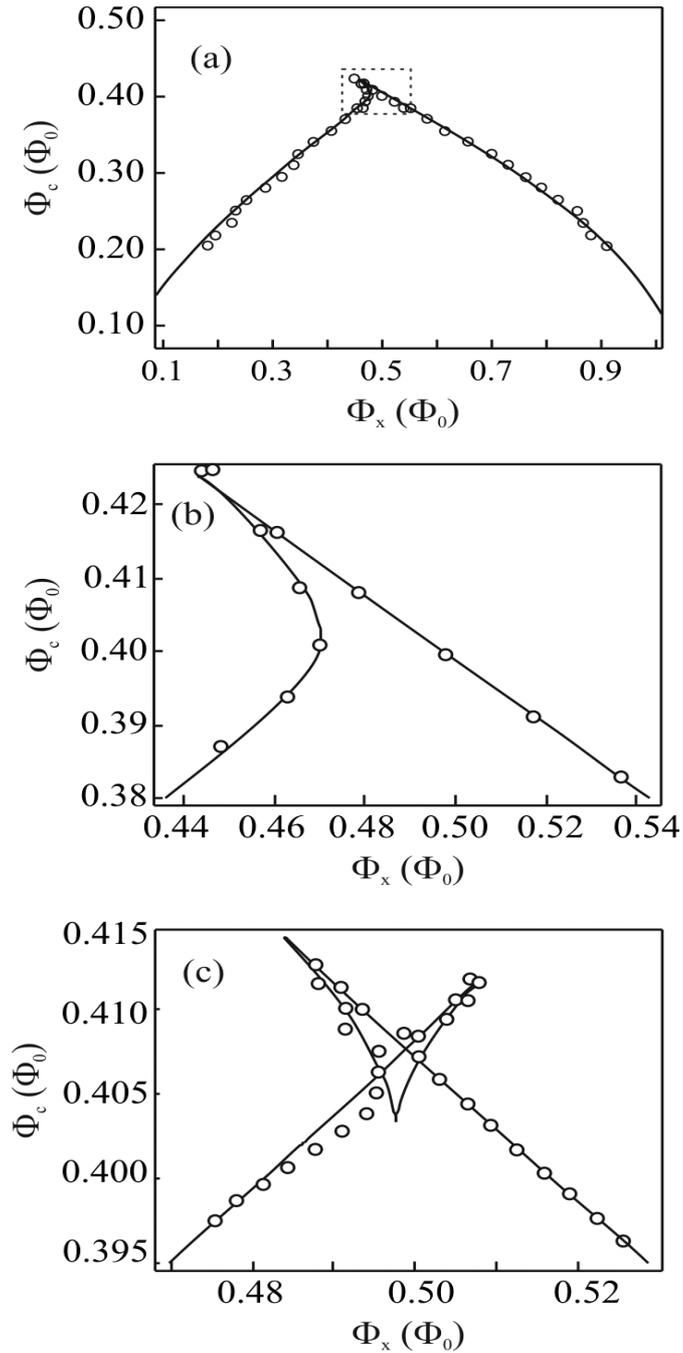

Figure 3. Experimental $\Phi_x$-$\Phi_c$ phase diagram obtained for two different asymmetry parameters determining the value of the coefficient to the third order power coefficient in eq. 3. (a) Difference between the critical currents is 4% ; (b) represents an enlargement of the tip region. Continuous line is the theoretical expression while the circles are the experimental points.(c) Butterfly catastrophe observed by enlarging the tip of the map when the coefficient to the third order in eq. 3 is close to zero (asymmetry of 0.3%).

The shape of the flux characteristics for $\Phi_c \cong \Phi_0/2$ can be fitted and provides information on the physical parameters of the inner dc-SQUID junctions. The data plotted in Fig. 3a indicate an asymmetry of about 4% between the critical currents of the junctions forming the inner junction loop. We attribute this asymmetry to flux trapped in the junctions; the validity of this hypotesis is confirmed by the fact that improving the shielding of the samples results in more symmetrical flux characteristics (we estimate an asymmetry between the critical currents of about 0.3%) and the backbending of Fig. 3a disappears. However, instead of the backbending we now observe "two-horns" on the tip shown in Fig. 3c. Unfortunately, both the asymmetrical pattern in Fig. 3a and the more symmetrical one in Fig. 3b are disturbing factors for our quantum coherence experiments because they indicate a modified shape of the potential in the lowest barrier region: in order to understand this phenomenon (and to possibly circumvent it) we put the observations into the following context.

The inner dc-SQUID contributes additional complexity to the potential energy function when the inductance $l$ of the inner loop of Fig. 1a cannot be neglected. With the phase differences between the quantum mechanical wave functions in the inner loop denoted by $\varphi_1$ and $\varphi_2$, it can be shown that the system can be described by the single variable $\varphi=(\varphi_1+\varphi_2)/2$ when the inner loop inductance is small compared to the total inductance. The approximate potential energy function reads[8]

$$U = \frac{1}{2}E_L(\varphi - \varphi_x)^2 - E_S \cos\varphi + E_D \sin\varphi - \frac{1}{2}E_l \cos^2\varphi \ , \qquad (1)$$

where $\Phi_0 = 2.07 \times 10^{-15}$ Wb is the magnetic flux quantum, $\varphi_x = 2\pi\,\Phi_x/\Phi_0$ and there are four energy scales. The first, $E_L = \Phi_0^2/(4\pi^2 L)$, is related to the magnetic energy stored in the main loop, while $E_S = E_{S0}\cos(\pi\,\Phi_c/\Phi_0)$, $E_D = E_{D0}\sin(\pi\,\Phi_c/\Phi_0)$ and $E_l = E_{l0} \sin^2(\pi\Phi_c / \Phi_0)$ represent respectively the harmonic modulations (through the applied flux $\Phi_c$) of $E_{S0} = (I_{C1} + I_{C2})\Phi_0 / (2\pi)$ (maximum Josephson energy of the two junctions of the internal loop), $E_{D0} = (I_{C1} - I_{C2})\Phi_0 / (2\pi)$ (Josephson energy due to the difference of the critical currents) and $E_{l0} = (I_{C1} + I_{C2})^2 l / 4$ (energy stored in the inductor $l$). Thus, eq. 1 describes the potential energy of the system by one variable, and four characteristic energies. The potential of eq. 1 can be derived following the analysis performed in ref. 8 for the two variables $\varphi=(\varphi_1+\varphi_2)/2$ and $\psi=(\varphi_1-\varphi_2)/2$ and bearing in mind that the network equations have to include the fact that the currents in inner and outer loop are nested. Starting from eqs. 1 and 2 of ref. 8 (with no noise current terms) but setting different Josephson currents ($I_{C1}$ and $I_{C2}$) for the junctions of the inner loop one derives the following equation (analogous to eq. 12 of ref. 8)

$$\psi_{tt} + \alpha\psi_t + \cos\varphi\sin\psi + \delta I_c \cos\psi\sin\varphi = -\frac{2}{\beta_l}(\psi - \pi\frac{\Phi_c}{\Phi_0})$$

where $\delta I_C = (I_{C1}-I_{C2})/2I_C$ , $I_C = (I_{C1}+I_{C2})/2$ and $\beta_l = (2\pi l\, I_C)/\Phi_0$. From this equation, following a linear expansion for the static limit performed for small $\beta_l$ and assuming a small $\delta I_C$, a single equation for the variable $\varphi$ (analogous of eq. 16 of ref. 8) and related potential energy (1) can be derived.

Given the values of inductances and currents in the actual experimental configuration, the four energies introduced above can be modulated by varying the two fluxes coupled to the system ($\Phi_x$ and $\Phi_c$). The difference between the Josephson currents of the junctions of the inner loop, related to the third term in eq. 1, can be due to either fabrication parameter uncertainties or flux trapping in the junctions or it can be imposed deliberately through the design of the chips. The values of currents and of inductances, however, are fixed for every measurement run and we can only probe the energy of the system through the externally applied fluxes and currents. It is natural then to display the stable and metastable states of the variable as a function of the control fluxes $\Phi_x$ and $\Phi_c$. A representative picture of this is shown in Fig. 4a, where the horizontal plane is spanned by $\Phi_x$ and $\Phi_c$ while the vertical axis is given by the phase variable $\Phi = (\Phi_0 / 2\pi) \varphi$. The surface shown in the figure is a known feature of Thom's Catastrophe theory [9,10,11] and, in the jargon of this theory, it is called the behaviour surface of the system; the surface is analytically determined by the fix-points of the system described by eq. 1; i.e., the zero point of the first derivative of eq. 1:

$$\varphi_x = \varphi + \frac{E_S}{E_L} \sin\varphi + \frac{E_D}{E_L} \cos\varphi + \frac{E_I}{E_L} \sin\varphi \cos\varphi \qquad (2)$$

The fold in the surface of Fig. 4a maps discontinuous transitions between multiple states that occur due to a change in a variable. The critical point at which a catastrophe occurs (i.e., a discontinuous transition from one point to another on a folded surface) corresponds to a point where a minimum and a maximum coincide. These values can be determined from the two conditions $\partial U / \partial \varphi = 0$ and $\partial^2 U / \partial \varphi^2 = 0$, which yield the critical value of $\varphi_x$. For fixed values of other parameters contributing to $E_S, E_D, E_I, E_L$ we thereby obtain plots of the critical points in the $\Phi_x$ - $\Phi_c$ plane, as shown in figures 4b-d.

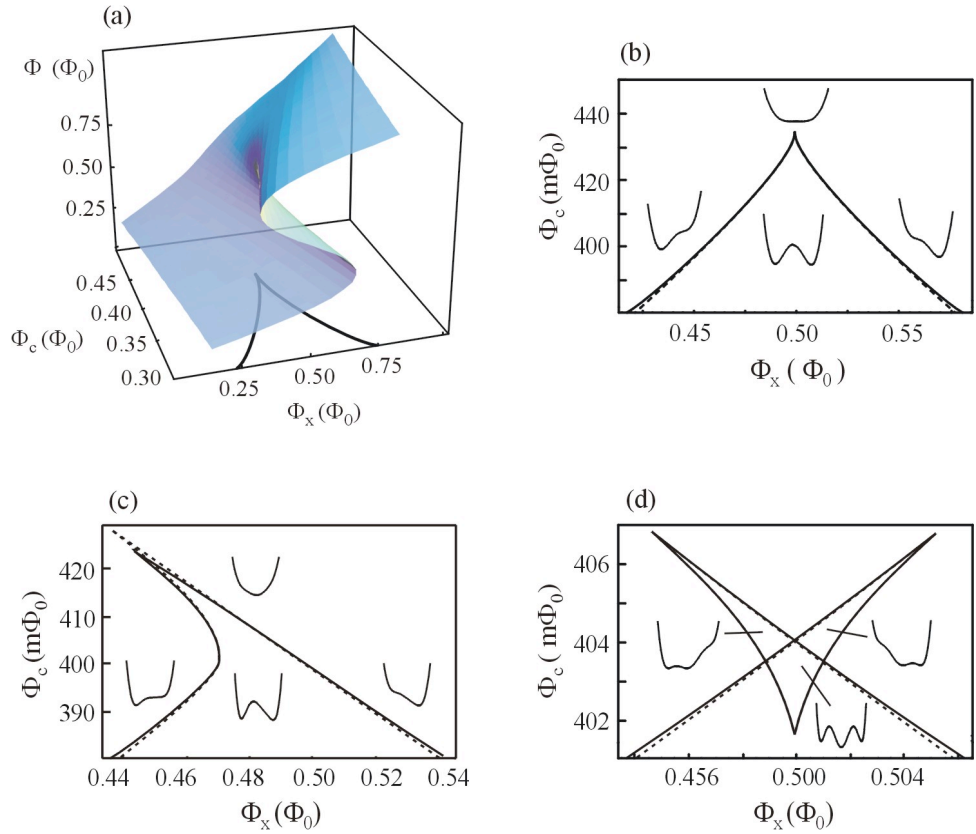

Figure 4. (a) Characteristic surface described by possible phase solutions of a double SQUID system. Shown surface is obtained by varying the two parameters that can be changed during measurements; this kind of surface represents a typical topological feature of systems exhibiting catastrophes. (b-d) Examples of parameter locations of catastrophes and sketch of the corresponding configurations of the potential (eq. 3). (b) Parameter values identical to (a); (c) and (d) are projections of the surface for different parameter sets of the butterfly catastrophe. The critical positions for which a catastrophe (a topological jump) occurs correspond to a point where a minimum and a maximum coincide. Solid and dashed curves are obtained by using respectively the full potential or a 6th order Taylor expansion.

Equation 1 can be readily transformed into the polynomial form analyzed by Thom [9-11] for the butterfly catastrophe by approximating the energy expression with a sixth order polynomial Taylor expansion around the point $\varphi = \pi$ (corresponding to $\Phi = \Phi_0/2$) ; removing the constant term and retaining only the leading power in the terms having the same coefficients we get

$$U = \frac{(\varphi - \pi)^6}{60} E_l - \frac{(\varphi - \pi)^4}{24}(4E_l - E_S) + \frac{(\varphi - \pi)^3}{6} E_D +$$
$$+ \frac{(\varphi - \pi)^2}{2}(E_L + E_l - E_S) - \varphi(E_D + \varphi_x - \pi) \qquad (3)$$

Thus, four physically relevant energies of our system directly determine the coefficients of the 6[th] order polynomial of the butterfly catastrophe. The phase diagrams shown in figures from 4b to 4d are related to three different sets of energy scales relevant for our experiment, and are obtained by the full potential in eq.1 (solid lines), and by the approximate potential in eq.3 (dashed lines): it is evident that the polynomial approximation is a very good fit of the original potential. We note that when the energy stored in the inductor $l$ can be neglected, and when there is no Josephson energy associated with the difference of the currents of the internal loop (the junctions are identical), the form (3) becomes a fourth order polynomial of the cusp catastrophe accounting for the behavior of the rf SQUID (a superconducting loop interrupted by a single Josephson junction). Equation 3 very clearly displays the relevance of the inductance $l$, since it determines the coefficient to the term of highest order, as well as the sign of the fourth order term.

The important parameters that we get out of the fit are $E_{S0}/E_L$, $E_{D0}/E_L$ and $E_{l0}/E_L$. From these quantities it is possible to determine all the system parameters $I_{C1}$, $I_{C2}$ and $l$ once the inductance $L$ is known (in our device we have a nominal inductance $L=85pH$). In Fig. 3a,b we show the theoretical curves fitting the data that were obtained with $E_{S0}/E_L = 4.9$, $E_{D0}/E_L = 0.196$ and $E_{l0}/E_L = 0.54$, corresponding to $l = 7.7\,pH$, $(I_{C1} + I_{C2}) = 19.0\,\mu A$, and asymmetry $\left(I_{C1} - I_{C2}\right)/\left(I_{C1} + I_{C2}\right) = E_{D0}/E_{S0} = 0.04$. These parameters are fully consistent with the asymmetry value derived from the flux modulations, meaning that the potential model of eq. 1 provides a realistic description of our system.

In Fig. 3c we further show that the experimental points fit the shape of the butterfly catastrophe that we obtain from eq. 2 for small junctions asymmetry (corresponding to a coefficient of the third order power close to zero in eq.3). The theoretical curve is the typical butterfly catastrophe and is obtained for $E_{S0}/E_L = 5.2$, $E_{D0}/E_L = 0.0156$ and $E_{I0}/E_L = 0.61$, corresponding to $l = 7.7\,pH$, $\left(I_{C1} + I_{C2}\right) = 20.1\,\mu A$, and $\left(I_{C1} - I_{C2}\right)/\left(I_{C1} + I_{C2}\right) = E_{D0}/E_{S0} = 0.003$. The asymmetry is consistent with the estimate obtained from the flux modulations. It is worth noting that no evidence of neither the tip bending shown in Fig. 3a,b and nor of the butterfly catastrophe shown in Fig. 3b was recorded at 4.2K; the data shown in these figures were recorded at 10 mK. The thermal fluctuations (not included in our model) dominate the effects of the finite inductance of the inner loop at 4.2K. In terms of potential wells (see Fig. 4d) the butterfly implies that an extra potential well exists between the right and left wells : when the control parameters lead the system in the "pocket" of the butterfly (see Fig. 4d) we can clearly see that the potential develops this third and central well in which the current is circulating only in the small loop of our double SQUID system.

From the coefficients of eq. 3 it is straightforward to anticipate the behavior that we observed experimentally. The coefficient of the fourth order term in that equation can be conveniently modulated experimentally via the normalized flux $\Phi_e/\Phi_0$. A particular feature of the experiments is that this coefficient is always negative, which is an important component for the butterfly catastrophe surface. Also, the coefficient regulating the shape of the tip is the third order term, namely $E_D$: when this term is close to zero we find experimentally the symmetrical projection of the butterfly singularity, just as predicted by the topological model [9-11]. The butterfly singularity has an important role in the catastrophe theory, and it has been invoked to explain psychological and social issues such as anorexia nervosa and war policy. In physics, an interesting analysis related to the butterfly catastrophe was reported for a three level optical system[12].

Due to the presence of the three-well structure in the sample considered in this paper, performing quantum coherence experiments according to our initial idea is a non-trivial task. However, simulation shows that, with a slight modification of the parameter values (for instance, a reduction of the critical current by about 15%), the third well disappears in the operational region and the overall modification of the potential profile is reduced to a small perturbation. Thus, an adequate design of the SQUID system should allow recovering the original measurement scheme.

In conclusion, our characterization of a double SQUID potential has shown that small deviations from ideal conditions have profound impacts on the shape of the potential energy. We have shown that the modification of the potential can be explained in terms of the general nonlinear system analysis introduced by R. Thom; namely catastrophe theory. The theoretical description of our experimental results obtained according to this model is accurate and consistent with independent parameter evaluations. We believe that quantum coherence experiments based on the Josephson flux variable can benefit from the analysis and the characterization of the potential that we have presented herein.

The chips used for the experiments, made in a standard Nb trilayer technology [13], were fabricated by Hypres Inc , NJ (USA). We wish to thank Prof. R. de Bruyn Ouboter for enthusiastic encouragement. The work was partially supported by INFN (Italy), by European Community Project RSFQubit (FP6-3748), by a MIUR-COFIN04 (Italy) and by a MIUR-FIRB03 (Italy). Partial support by AFOSR grant FA9550-04-1-0711 through the Center for Digital Security, UC Davis (USA) is also acknowledged.